# Strategies for spectroscopy on Extremely Large Telescopes: 1 – Image Slicing

J.R. Allington-Smith[*]

Centre for Advanced Instrumentation

Physics Dept, Durham University, South Rd, Durham DH1 3LE

j.r.allington-smith@durham.ac.uk

**Abstract**

One of the problems of producing spectrographs for Extremely Large Telescopes (ELTs) is that the beam size is required to scale with telescope aperture if all other parameters are held constant, leading to enormous size and implied cost. This is a particular problem for image sizes much larger than the diffraction limit, as is likely to be the case if Adaptive Optics systems are not able initially to deliver highly-corrected images over the full field of the instrument or if signal/noise considerations require large spatial samples. In this case, there is a potential advantage in image slicing to reduce the effective slit width and hence the beam size. However this implies larger detectors and oversizing of the optics which may cancel out the advantage. By the means of a toy model of a spectrograph whose dimensions are calibrated using existing instruments, the size and relative cost of spectrographs for ELTs have been estimated. Using a range of scaling laws derived from the reference instruments it is possible to estimate the uncertainties in the predictions and to explore the consequences of different design strategies. The model predicts major cost savings (2-100×) by slicing with factors of 5-20 depending on the type of spectrograph. The predictions suggest that it is better to accommodate the multiplicity of slices within a single spectrograph rather than distribute them among smaller, cheaper replicas in a parallel architecture, but the replication option provides an attractive upgrade path to integral field spectroscopy (IFS) as the input image quality is improved. Another major issue is whether the camera speed should be adapted to minimise the cost of the detector, or conversely, the camera simplified by means of redundant oversampling which requires larger detector formats. This tradeoff is the main reason for the minimum in the size and cost predictions for moderate slicing.

## 1. Introduction: Quasi-diffraction-limited slicing

This is the first part of a study of strategies to produce instruments for Extremely Large Telescopes (ELTs) that are both affordable and usable before the full benefits of Adaptive Optics delivering near-diffraction-limited imaging become available.

It has long been realised that ELT instruments matched to natural seeing (0.3-0.7 arcsec FWHM depending on conditions and wavelength) must be very large and equipped with unfeasibly fast cameras. This is a simple consequence of the fact that the characteristic disperser dimension (in Littrow configurations, simply the length along the optical axis) must scale with telescope diameter for fixed resolving power if the slitwidth is constant; and the conservation of Etendue in classical optical systems. To defeat the former problem requires a reduction in the slitwidth since this allows the disperser dimension (and the beam diameter for a fixed blaze angle) to be reduced while delivering the same resolving power. The problem is that reducing the slitwidth also reduces the throughput by a large factor (if correctly oversampled). One solution is to exploit recent improvements in integral field unit (IFU) construction (e.g Dubbeldam et al. 1994, Allington-Smith et al. 1996a,b) to slice the input image efficiently into numerous thin slices that can be reformatted into a single long pseudo-slit. This requires extra detector pixels to accommodate the elongated slit and larger optics to accommodate the increased field angles.

---

[*] Email address: j.r.allington-smith@durham.ac.uk

This approach also opens up the possibility of directing the light in each slice to a separate spectrograph which could be of simple, compact construction and thus easy to replicate. This has synergy with instrument concepts such as MOMSI (Evans et al. 2006) and MOMFIS (Cuby et al. 2006) which divide the field into separate pieces, each of which is processed by an independent IFU. The same multiplexed unit spectrographs can thus be used whether it is the field or the slit that is sliced. Furthermore such an approach allows an orderly transition between dealing with natural seeing, in which case there is no spatial information to be recovered, and improvements in image quality as the AO systems mature, allowing spatial information to be recovered. From the point of view of the spectrograph, it is simply a question of whether you sum over all the pixels along the slit or attempt to recover spatial information from the same pixels: different software, same hardware.

In this paper, the potential benefits of slicing are examined by a simple model, described in §2, that attempts to account for all the main optical effects which might prevent the advantages being realised in practice. This is coupled with a costing model, in the spirit of those long used in industry, to estimate the costs of instruments of characteristic types on ELTs. This is based on a calibration using current 8-m telescopes of broadly similar characteristics to the ELT concepts in an attempt to reduce the uncertainty in this considerable extrapolation. This is presented in §3. In §4, results are obtained to determine the best strategy for designing instruments of this type for a 40-m telescope such as the proposed European ELT.

## 2. Scaling laws for instrument size

To explore the potential advantages of slicing, we need to construct a simple model of a spectrograph. The simplest example is an echelle spectrograph used in Littrow configuration to effect a double-pass system in which the spectrum is projected back onto the slit plane. A modification to this, which allows for a camera of different speed to that of the collimator will be described later.

Of course, most current instruments have quite different designs with greater complexity. Even high-resolution echelle spectrographs, on which the toy model is based, all use reflective collimators rather than a single transmissive optic which would be very difficult to manufacture. Furthermore this type of spectrograph normally has a separate camera (frequently transmissive) and substantial optics for cross-dispersion – which is ignored here, as are the slicing optics. However, it seems reasonable to expect that such a design would follow roughly the same scaling laws as the toy model. Whatever the details, it is important to use the simplest possible model to reduce the number of adjustable parameters and to be representative of the highly space-efficient designs which will be mandatory for ELTs. Finally, even if this model is too simple, it can still be used as a design tool for the design of more realistic instruments by adjusting the scaling parameters and/or adding complexity[†]. This will allow an exploration of the interrelationship between the design of the instruments and the telescope to ameliorate the large financial risk – something which is far more important for ELTs than it ever was for 8-10m telescopes.

In the model, the slit is either illuminated directly by the telescope (focal ratio $F_T$) or takes the form of a pseudo-slit comprising an array of optical fibres or slicing optics relaying light from the telescope focus. This slit may be sliced into $N$ pieces which are arranged end to end to form the pseudo-slit that is both narrower and longer than the original by the factor $N$.

For a negligible field, with monochromatic light, the spectrograph size can be estimated from the beam diameter, $D$, the input focal ratio, $F_T$, and the grating blaze angle, $\gamma$, alone. However the spectrograph/camera optic needs to be enlarged to accommodate light whose principal ray angle is $\pm\Psi/2$ due to off-axis points in the field and $\pm\Phi/2$ due to the simultaneous range of wavelengths in the spectrum. This is illustrated in the top panels of Fig.1. The dimensions of the spectrograph may then be

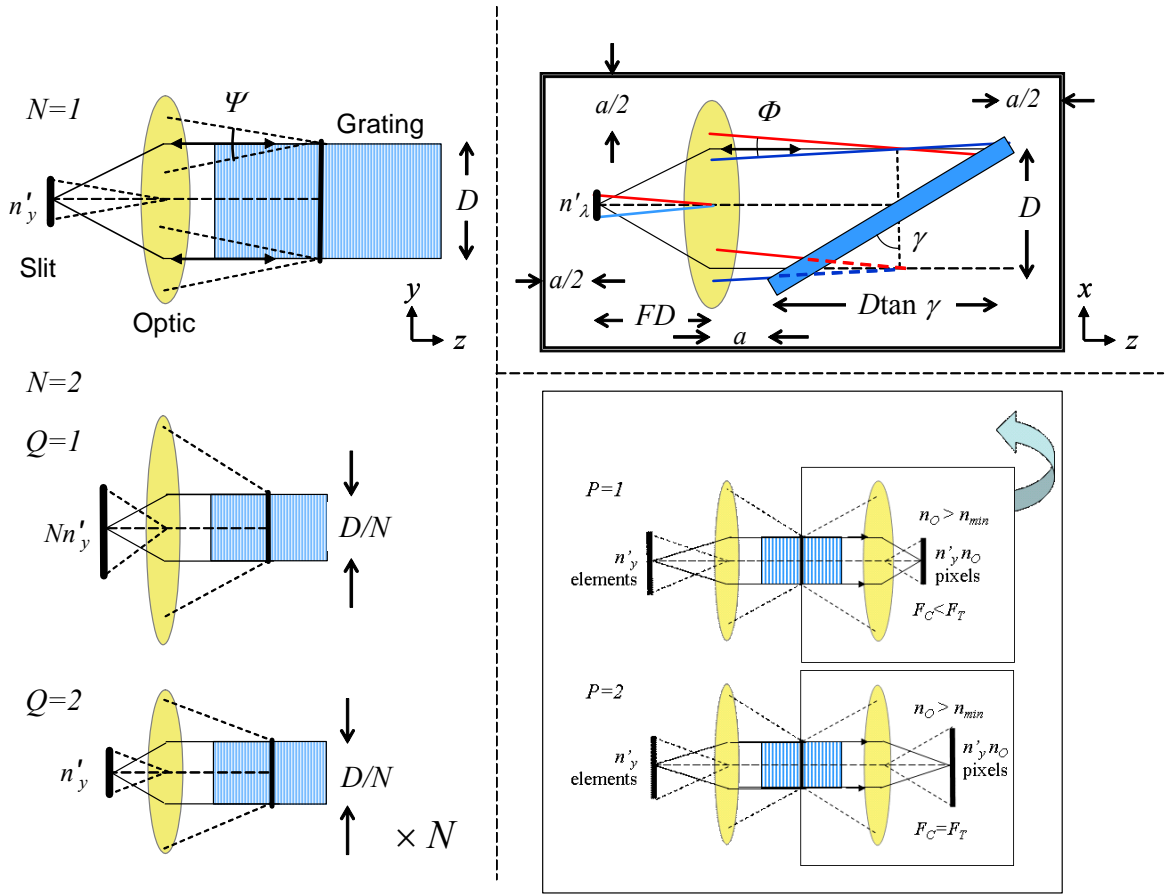

**Figure 1: Schematic of spectrograph model described in text. The effect of slicing is shown for slicing factor $N=2$ for the two cases described in §2, labelled $Q=1$ and $Q=2$. For $Q=2$, the spectrograph is replicated $N$ times. For the x-z projection, the oversize factor $a$ is indicated. The inset at bottom-right illustrates the difference between the $P=1$ and $P=2$ models described in §2. For convenience the spectral rays in the x-z projection are drawn as converging on the implied pupil stop.**

estimated as

$$L_x = \{D + \Phi(a + D\tan\gamma) + a\}b$$
$$L_y = \{D + \Psi(a + D\tan\gamma) + a\}b \qquad (1)$$
$$L_z = \{D(\max(F_T, F_C)) + D\tan\gamma + 2a\}b$$

$F_C$ is the camera focal ratio which is allowed to take independent values from the collimator focal ratio determined by the telescope, $F_T$. In the spirit of retaining a compacted, quasi-double-pass construction, the slower of these determines the spectrograph length. It has been assumed that the full grating depth along-axis contributes to the required oversizing despite the convergence of collimated beams from different parts of the field at the image of the telescope exit pupil half-way along the disperser's depth. The true situation is rather more complex and requires proper ray-tracing, but using the full-depth is a reasonably-conservative approximation. To make the model more realistic, we assume that the optics are oversized by an amount $a$ in $x$ and $y$. For convenience we use the same oversizing dimension in the $z$-direction to account for detectors and other equipment and also the clearance between the collimator optic and the disperser. Finally we multiply all dimensions by a factor $b$.

From the standard expression for the resolving power of a spectrograph in the Littrow configuration at the blaze condition (e.g. Lee and Allington-Smith 2000), $R = \lambda/\delta\lambda$, where $\delta\lambda$ is the spectral resolution, we obtain the beam diameter, $D$, in terms of $R$, the angular slit width, $\chi$, and the blaze angle, $\gamma$,

$$R = \frac{2\tan\gamma}{\chi D_T} D \;\Rightarrow\; D = \frac{\chi R D_T}{2\tan\gamma} \qquad (2)$$

The spectral beamspread is

$$\Phi = n'_\lambda \left(\frac{d\beta}{d\lambda}\right)\frac{\lambda}{R} \qquad (3)$$

Where $n'_\lambda$ is the number of resolution elements in the spectrum, each covering wavelength range $\delta\lambda = \lambda/R$. The angular dispersion $d\beta/d\lambda = m\rho/\cos\beta$ is obtained by differentiating the grating equation $\sin\alpha + \sin\beta = m\rho\lambda$ where $\alpha$ and $\beta$ are the angles on incidence and diffraction relative to the normal to the surface of a diffraction grating with ruling density, $\rho$ operating in order $m$. At blaze $\alpha = \beta = \gamma$, so, substituting for $R$ from Eqn. (2),

$$\Phi = n'_\lambda \left[\frac{\lambda}{R}\frac{m\rho}{\cos\beta}\right]_{\beta=\gamma} = n'_\lambda \frac{2\tan\gamma}{R} = n'_\lambda \chi \frac{D_T}{D} \qquad (4)$$

Note that $\chi(D_T/D)$ is the angular spread from one slitwidth due to conservation of Etendue. The spatial beamspread is obtained directly from Etendue conservation in terms of the angular slit length, $\xi$, as

$$\Psi = \xi\frac{D_T}{D} = n'_y \chi \frac{D_T}{D} = n'_y \frac{2\tan\gamma}{R} \qquad (5)$$

using Eqn.. (2) and since $\xi = n'_y\chi$ assuming isotropic oversampling. Substituting for $D$, $\Phi$ and $\Psi$ from (2)-(5) into (1) then gives

$$L_x(\chi, n'_\lambda) = \left\{ \chi D_T \left( \frac{R}{2\tan\gamma} + n'_\lambda \tan\gamma \right) + a\left(1 + n'_\lambda \frac{2\tan\gamma}{R}\right) \right\} b$$

$$L_y(\chi, n'_y) = \left\{ \chi D_T \left( \frac{R}{2\tan\gamma} + n'_y \tan\gamma \right) + a\left(1 + n'_y \frac{2\tan\gamma}{R}\right) \right\} b \qquad (6)$$

$$L_z(\chi, F_T, F_C) = \left\{ \chi D_T \left( \frac{R}{2\tan\gamma} [\max(F_T, F_C) + \tan\gamma] \right) + 2a \right\} b$$

Note that the dimensions depend only on the angular slitwidth if the resolving power and blaze of the grating are held constant for a given design of spectrograph and telescope.

When the slit is sliced by factor $N$, the only thing to be changed in Eqn. (6) is the angular slitwidth which scales as $\chi(N) = \chi_1/N$ where $\chi_1 = \chi(N=1)$. This reduces the size of the spectrograph stop and beam diameter resulting in a more compact instrument. Slicing may also implicitly affect the camera focal ratio and/or oversampling of the slit image by the detector which will affect the number of detector pixels required to provide the same number of spatial resoultion elements.

There are two strategies for preserving the information content $n'_\lambda n'_y$, which are subject to limitations on both the speed of the camera and the minimum useful sampling (summarised in Figure 2). The camera focal ratio and the oversampling by the detector, $n_O$, are linked by the physical size of the image on the detector surface, $\chi F_C D_T = n_O d_p$ where $d_p$ is the physical size of each square pixel (typically $13 < d_p < 28$ µm depending on detector type).

***P = 1:*** Adjust the camera focal ratio to give a fixed oversampling of $n_O = n_F$. Although the value of $n_F$ is arbitrary, it is constrained to exceed a value $n_{min} = 2.5$, slightly greater than the requirement for Nyquist sampling, allowing for the non-gaussian shape of the point-spread function. The camera focal length is constrained to $F_C > F_{min} \approx 1.5$, the fastest camera design available. Thus

$$F_C(N,1) = N n_F \left( \frac{d_p}{\chi_1 D_T} \right) > F_{min} \qquad n_O(N,1) = \frac{F_{min}}{N}\left( \frac{\chi_1 D_T}{d_p} \right) > n_F \quad (7)$$

where $F_C(N,1)$ is the camera focal ratio for slicing factor, $N$, for model $P = 1$. This option is the most economical in terms of numbers of detector pixels but requires a longer camera for large slicing factors.

***P = 2***: Keep the camera focal ratio the same as the collimator, to facilitate a compact double-pass construction, and adjust the oversampling as necessary. Thus

$$F_C(N,2) = F_T > F_{min} \qquad n_O(N,2) = \frac{F_T}{N}\left( \frac{\chi_1 D_T}{d_p} \right) > n_{min} \qquad (8)$$

where $n_{min}$ and $F_{min}$ are the same oversampling and focal ratio limits decribed above. This buys simplicity and compactness at the expense of an increase in the number of detector pixels which is proportion to the square of $n_O(N)$.

Finally, the extra pixels required for the $N$ slices must be accommodated. There are two options labelled by $Q$.

*Q = 1:* Add them to the length of the slit, in which case we require a further *N*-fold extension to the number of detector pixels in the spatial direction.

*Q = 2:* Direct the slices into *N* replicas of the original spectrograph in which case we need *N* separate spectrographs but keep the same number of spatial resolution elements in each spectrograph.

In terms of *Q*, the number of spectrographs scales with $N^{Q-1}$ and the number of spatial resolution elements in *y* scales as $N^{2-Q}$. We can therefore estimate the total instrument volume as

$$V(P,Q,N) = N^{Q-1} L_x\left(\frac{\chi_1}{N}, n'_\lambda\right) L_y\left(\frac{\chi_1}{N}, N^{2-Q} n'_y\right) L_z\left(\frac{\chi_1}{N}, F_C(N,P)\right) \quad (9)$$

Where $F_C(N,P)$ is given by equations (7) or (8). Note that the instrument volume is independent of the numbers of pixels in the detector array which scales with the oversampling factor according to eqn (7) or (8). However, this will affect the cost. In the results shown below, we plot instrument size, estimated as the cube-root of the total volume, rather than the volume itself, since the former is easier to understand intuitively.

## 3.  Cost model and calibration

The cost is estimated by scaling the non-detector part by the instrument volume and the detector sub-system by the number of pixels in total

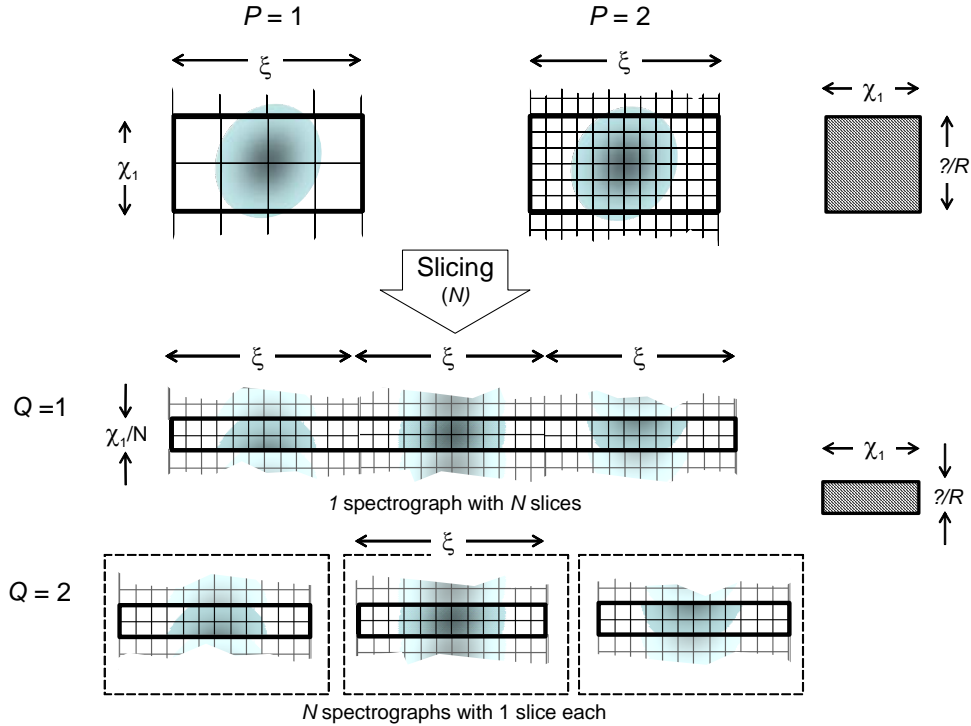

**Figure 2: Summary of the slicing process to illustrate the two pairs of mode options labelled by P and Q. The shaded box indicates the size of the resolution element. In the example shown, $N = 3$, $n_F = 2$ and $F_C(N) = F_T$.**

$$C(P,Q,N) = \frac{V(P,Q,N)}{V_R}(C_{TR} - C_{DR}) + \left(\frac{Nn'_x n'_y n_O^2(N,P)}{n_{xR} n_{yR}}\right) C_{DR} \qquad (10)$$

Where $V_R$ is the volume of a reference spectrograph of cost $C_{TR}$ where $C_{DR}$ is the cost of the detector subsystem containing $n_{xR}, n_{yR}$ pixels. The assumption that cost scales with instrument volume is widely adopted in the aerospace industry for space missions and some astronomical instrumentation institutes, but with *ad hoc* modifications depending on the special features of the instrument (e.g. X-ray or visible/IR; passively or actively-cooled). The equivalent of that adjustment in this context is to allow for the difference in cost of visible and infrared detectors. There is no justification for allowing the instrument cost to scale with volume with an arbitrary exponent since the dimensions are fixed by simple optical considerations and the linear relationship between volume and cost is both widely accepted and intuitively reasonable given that the mass of both the active optics and the structures required to support them are likely to scale with volume.

However, one could make a case that the cost will have a super-linear dependency on volume given that the stability requirement is set by the physical detector pixel size which is independent of telescope diameter, but the mass of the components which need to be kept stable scales with telescope size to some power (from §4 instrument volume and assumed mass $\propto D_T^\kappa$ with $1 < \kappa < 2$). For a cantilever of elastic modulus $E$ and length $L$ carrying mass $M$, the end deflection is $\propto ML^3/E$. Thus for fixed deflection, we require an increased stiffness of the structure of $E \propto D_T^\eta$ where $\kappa < \eta < 2\kappa$, i.e. a factor 10–300 between telescope apertures of 8 and 40m depending on whether the supported component scales with the telescope aperture or not. This reinforces the widely-held view that it will be necessary to mount the instruments on gravitationally-stable platforms and use active flexure control for the optical pickoff attached to the telescope[‡]. Even so, it seems likely that extra mass may be required to stiffen the structure. Therefore, the assumption that cost scales linearly with instrument volume seems to be rather conservative.

The scaling parameters *a* and *b* are adjusted until the model volume equals the actual volume using the three representative calibration examples shown in Table 1, These are all facility instruments of the Gemini observatory whose costs are on a relatively uniform basis for comparison and represent relatively good value for money: GMOS (Hook et al. 2003), GNIRS (Elias et al. 2006) and bHROS (Aderin 2004). The exact costs are not in the public domain but we assume $C_{TR}$ = 5M (we use US$ as the

**Table 1: Summary of instruments used for calibration with the two sets of scaling parameters which fit the estimated actual volume to within 10% accuracy.**

|  | $\chi_1$ (″) | $N$ | $R$ | $\lambda$ (nm) | $\gamma$ (°) | $n_\lambda$ | $n_y$ | $n_O$ | $D$ (mm) | $V_R$ (m$^3$) | $C_{DR}$ (M$) | $S$ | $a$ (mm) | $b$ |
|---|---|---|---|---|---|---|---|---|---|---|---|---|---|---|
| GMOS | 0.5 | 1 | 5000 | 400 | 26 | 6144 | 4608 | 6 | 100 | 2 | 0.25 | 1 | 100 | 2.7 |
|  |  |  |  |  |  |  |  |  |  |  |  | 2 | 520 | 1.1 |
| GNIRS | 0.3 | 1 | 5900 | 2140 | 13 | 1024 | 1024 | 2 | 150 | 2 | 0.5 | 1 | 100 | 2.1 |
|  |  |  |  |  |  |  |  |  |  |  |  | 2 | 460 | 1.1 |
| bHROS | 0.7 | 5 | 150000 | 400 | 63 | 2048 | 4096 | 3 | 210 | 5 | 0.25 | 1 | 100 | 1.5 |
|  |  |  |  |  |  |  |  |  |  |  |  | 2 | 330 | 1.1 |

currency unit throughout) for each. Costs for the detector systems are based on commercially available units (controller plus camera head).

Of course, there is no unique solution for the scaling parameters so two extremal sets were obtained on the following assumptions, labelled by *S*.

**S = 1:** the scaling is dominated by the proportional factor, *b*, so we set the fixed oversize factor to be small, *a* = 100mm and then find *b*.

**S = 2;** the scaling is dominated by the fixed amount, *a*, so we set *b*= 1.1 and then find solutions for *a*.

From Table 1, we see that the scaling parameters vary considerably with a range of 30% in *a* and 60% in b. This suggests that we should not adopt average parameter values but use the parameter set for the reference instrument that most closely resembles the instrument concept under investigation. Note that the tacit assumption is made that the ELT instruments will not be more space-efficient than current ones.

GMOS

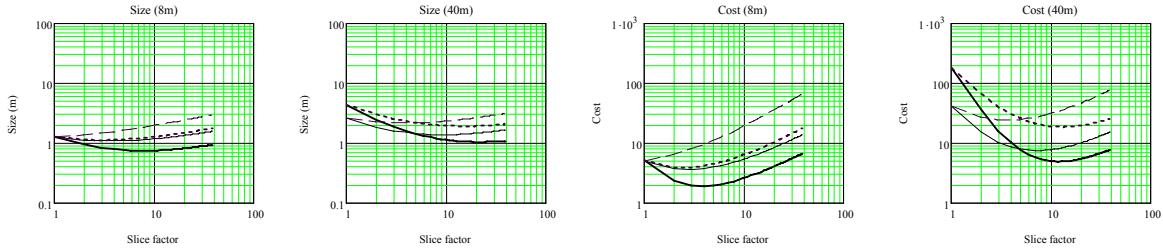

GNIRS

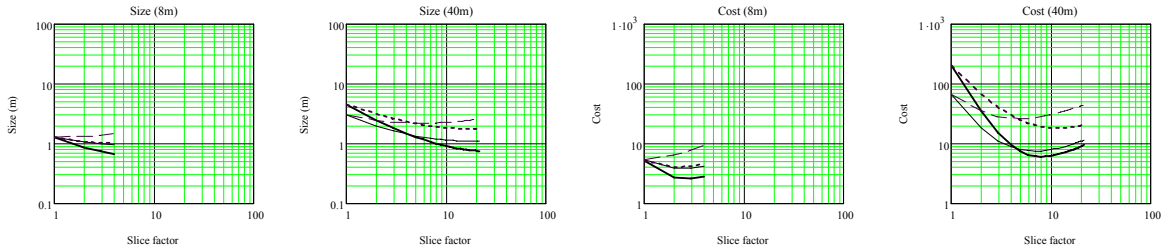

bHROS

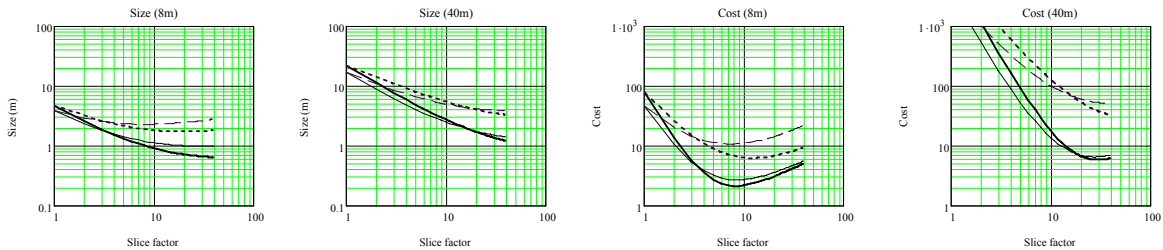

**Figure 3: Size (left pair) and estimated cost (right pair) of three Gemini instruments based on the model discussed in the text for an 8m telescope (left of each pair) and a 40-m telescope (right of each pair). Key to curves: heavy solid – Q=1 for scaling parameters S=1, Light solid – Q=1 for S=2; Heavy dashed – Q=2 for S= 1; Light dashed - Q=2 for S=2. The curves terminate when the diffraction limit is reached or *N* = 40. The unit of cost is millions of US$.**

The oversizing parameters are highest for GMOS, (e.g. $b$ =2.8). This may be due to any or all of the following considerations. (a) It is an uncooled instrument where compactness is less essential than for an instrument such as GNIRS where the entire optical train must fit within a cryostat. (b) It is optimised for a wide field which requires a complicated, hence bulky, optical train (6 groups of triplet lenses); (c) It was intended to be a more versatile instrument than the others (the only one with identical copies on the two Gemini telescopes) which implies additional complexity. bHROS has the smallest oversizing parameters but is the most specialised of the three and conforms most closely to the toy-model since it uses an echelle grating.

A scaling law for detector cost versus numbers of pixels is also required. It is assumed that detector cost is proportional to a power, $v$, of the relative number of pixels and made estimates using $v = 0.8$, but will explore the effect of changing this parameter later.

## 4. Results

Fig. 3 shows cost and size predictions for the three reference instruments as listed in Table 1. Although the costs are given in absolute units of millions of US$, it is best to examine the *relative* costs of the strategies explored. The predictions for an 8m telescope provide a good match to the actual instruments in the unsliced case and for $N = 5$ in the case of bHROS for the $Q = 1$ case ($Q = 2$ is inappropriate since there is

only a single spectrograph). These results indicate what would happen if close analogues of these existing instruments were simply scaled to a 40-m telescope. Comparing the $N = 1$ values, it can be seen that there is an implied 5-30 times increase in cost leading to an instrument suite as expensive as the telescope! This is a powerful motivation to consider slicing as a means to reduce instrument size and cost. Slicing at the diffraction limit appears to be a poor strategy since the fixed oversizing parameter, *a,* and the oversizing due to the beamspread starts to dominates the instrument volume with the result that the size/cost actually increases beyond that of an unsliced instrument.

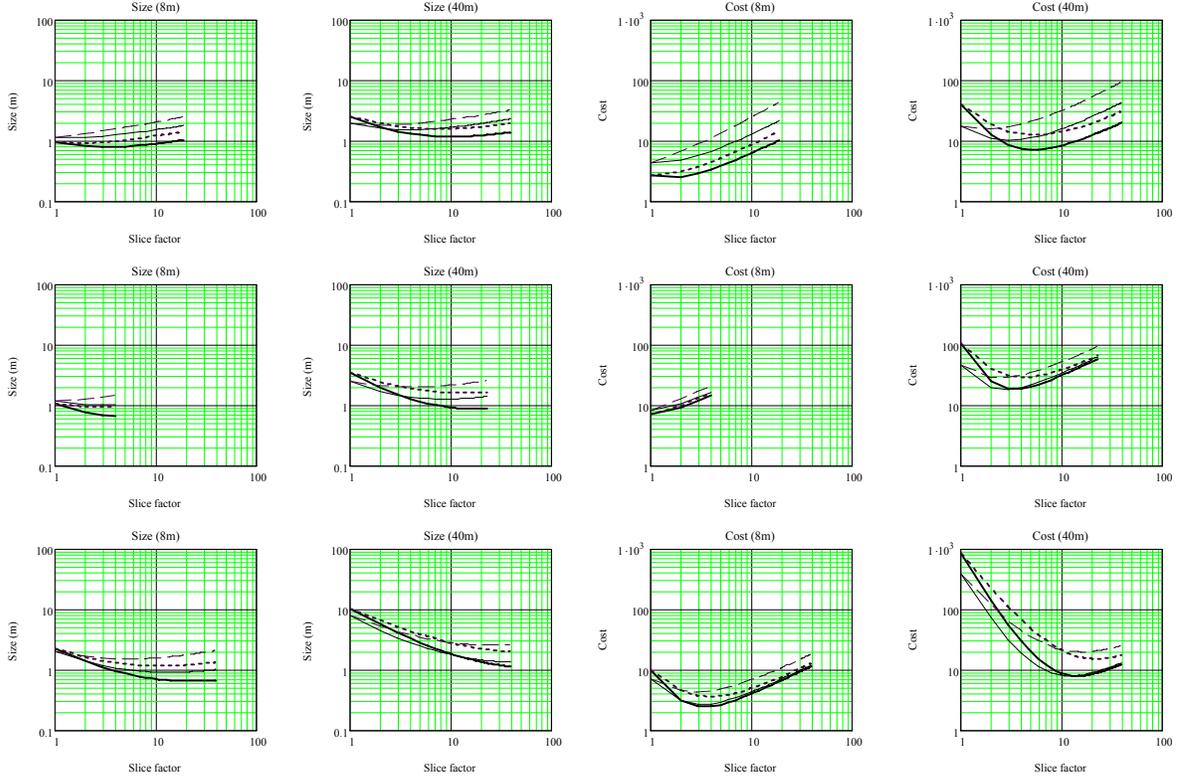

**Figure 4: Predicted cost and size for spectrographs using model $P$=1 as described in the text for the three cases summarised in Table 2 in the same order. Each curve is for different values of $Q$ and $S$ as described in the caption of Fig.3.**

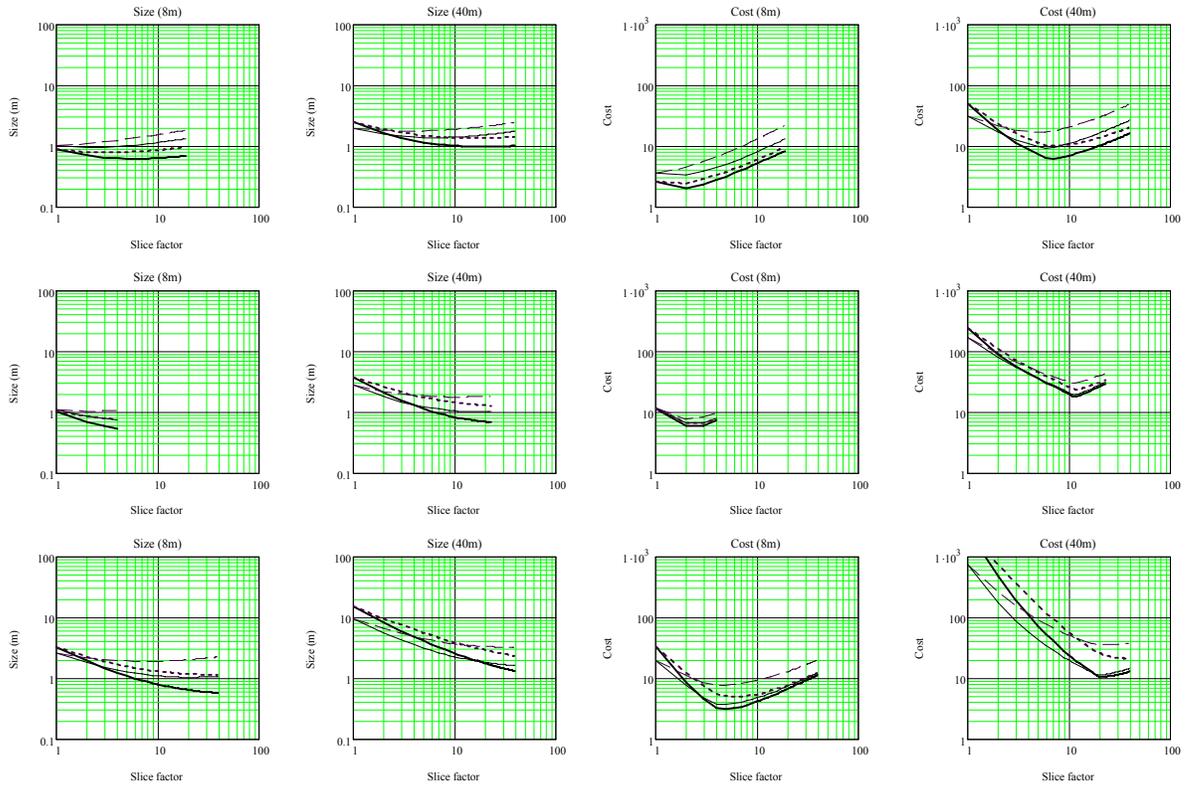

**Figure 5: Same as Fig. 4 but for *P* = 2.**

Another clear conclusion so far is that the $Q = 1$ model, in which the spectrographs are not replicated, almost always is the best strategy for reducing size and cost, by a substantial margin.

As noted above, the instrument volume scales with telescope aperture (for the same slicing factor) with an exponent in the range $1 < \kappa < 2$. This is much less severe than the naïve prediction of $\kappa = 3$. The reason for this is that the beam size is only one of the factors that contributes to the instrument volume (scaling as $\kappa = 3$) whereas the other factors have a weaker dependence on telescope aperture.

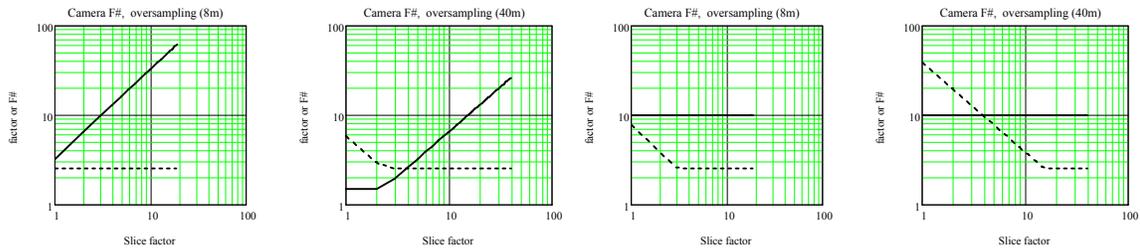

**Figure 6; Camera focal ratio (solid) and oversampling (dotted) for P=1 model (left pair) and P = 2 (right pair). For the case 1 (R = 5000, visible/red spectrograph) model.**

**Table 2: Summary of parameters used for predictions of cost and size.**

| Case | $\chi$ ('') | $R$ | $\lambda$ (nm) | $\gamma$ (°) | $n'_\lambda$ | $n'_y$ | $F_T$ | Cost ref | $P$ | $N_{opt}$ |
|---|---|---|---|---|---|---|---|---|---|---|
| 1  VisR-MOS | 0.4 | 5000 | 650 | 39 | 1640 | 1640 | 10 | GMOS | 1 | 5-6 |
|  |  |  |  |  |  |  |  |  | 2 | 6-8 |
| 2 NIR-MOS | 0.3 | 5000 | 2000 | 13 | 1640 | 1640 | 10 | GNIRS | 1 | 4-5 |
|  |  |  |  |  |  |  |  |  | 2 | 10 |
| 3: VisB-HRS | 0.6 | $1.5 \times 10^5$ | 400 | 76 | 1640 | 1640 | 10 | bHROS | 1 | 8-20 |
|  |  |  |  |  |  |  |  |  | 2 | 20 |

Figs 4 and 5 show predictions for three derivatives of the reference spectrographs with parameters more representative of current requirements for ELT instruments: 1 – a visible/red spectrograph; 2 – a near-infrared spectrograph, both with medium spectral resolution and suitable for mulitplexed spectroscopy; and 3 –  a high-resolution (single-object) spectrograph (see Table 2 for details). Results for both the $P = 1$ and $P = 2$ cases are shown.  Because these represent an extra extrapolation from the model calibration baseline, the predictions are more tentative than those discussed above. In particular, the relative costs of the three instrument concepts are likely to be especially inaccurate. However the cost versus slicing factor for a given instrument concept is still likely to be useful.

It is immediately clear from these figures that slicing has the potential to hugely reduce the size and cost of instruments with  savings of factors of at least 3 and up to ~100 with the biggest benefit to high resolution spectroscopy.

 Comparing the $P = 1$ and $P = 2$ models, it can be seen that this has a significant effect for large slicing factors in that the optimum slicing factor is greater for  $P = 2$, for which the camera speed is constrained to that of the collimator by allowing large oversampling factors at small slicing factors. This behaviour can be examined in more detail in Fig 6 which shows the variation of oversampling and camera speed for the 8-m and 40-m cases. For $P = 2$, for which the camera speed is fixed at $F_C = F_T = 10$, the oversampling reaches 40 in the unsliced case with a corresponding extra burden in cost. Thus, as the slicing factor increases, the cost of the detector sub-system, and the extra spectrograph volume to accommodate all those extra pixels, is reduced, thereby producing a cost minimum at larger slicing factor than in the $P = 1$ model. In this case, the numbers of pixels is relatively constant, except for low slicing factor where a modest increase in oversampling is required to avoid the need for an impracticably fast camera and at large slicing factors, the length of the camera becomes a significant contributor to the instrument volume. The optimum slicing factor is given in the last column of Table 2 for the different *P*-models.

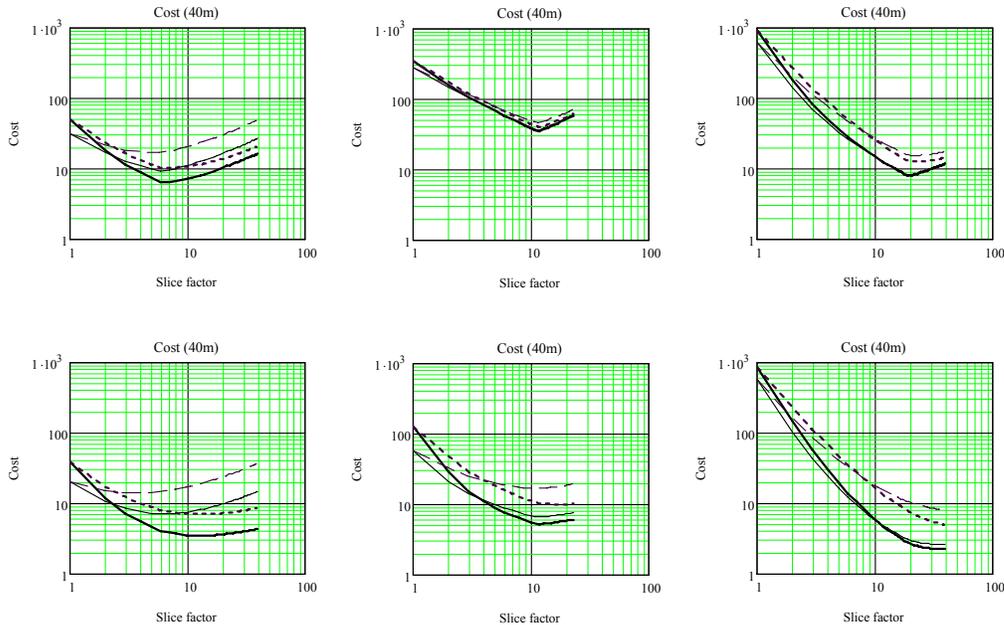

**Figure 7: Comparison of the effect of varying the detector cost scaling parameter, ν, for the three instrument cases 1 - 3 (left to right). Top row: ν = 0.8. Bottom row:  ν = 0.4**

The effect of changing the cost scaling parameter, ν, is shown in Fig.7. Clearly, making the detector pixel scaling law less severe produces very substantial reductions in cost for large slicing factors where the fractional cost of the detector dominates. This is particularly true for the NIR-MOS (case 2) concept which is scaled from GNIRS which has a relatively small detector format of 1024×1024. Because of this considerable extrapolation the difference in cost (a factor of ~8) should be interpreted with caution. In contrast, the effect on the VisR-MOS (case 1) based on the well-pixelled GMOS is much less dramatic. However the basic message that modest slicing dramatically reduces cost still holds and is even more convincing if one adopts a benign detector cost scaling law. Since, we may expect detector costs to reduce in real terms this assumption seems quite reasonable.

The effect of changing the telescope focal ratio from F/10 to F/16 is shown in Fig. 8. This has little effect although instrument volumes and cost are slightly increased.

From this, it appears that the best strategy to build a cheap spectrograph for a 40m

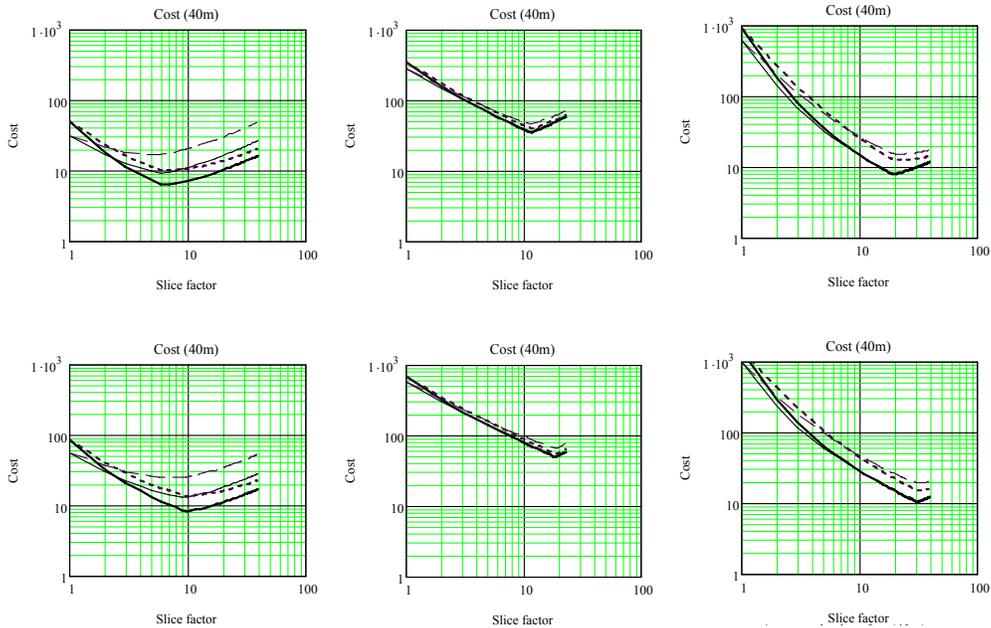

**Figure 8: Comparison of the effect of varying the telescope focal ratio from 10 (top) to 16 (bottom) for the 3 instrument cases 1-3 (left to right).**

telescope is as follows:

For a low-medium resolution spectrograph in the visible-near-IR spectrum, slice the slit by a factor of 5-8. Adapt the camera focal ratio to minimise the required oversampling and thus the number of pixels. Alternatively, similar cost could be obtained by slicing more and fixing the camera speed to that of the collimator. However, as with all conclusions reached here, where there is a choice of slicing factor, it is best to choose a small value because the cost of the slicing optics is not included in the model; so an unnecessarily large slicing factor may result in extra hidden costs. However if the spectrograph is intended for integral field spectroscopy, once the AO system is able to provide sufficiently good image quality, large slicing factors will be required anyhow which favours the $P = 2$ model. The results presented here imply a small reduction in cost for a useful $N = 20$ slicing factor compared to the unsliced case, if $P = 2$; i.e. the oversampling is adapted to a fixed, slow camera.

For a high-resolution (blue/visible) spectrograph, a higher slicing factor of 10-20 is recommended. Once again, the $P = 1$ model is favoured on the grounds of reduction in risk.

## 5. Conclusions

This is the first part of a study of strategies to produce instruments for Extremely Large Telescopes that are both affordable and usable before the full benefits of Adaptive Optics delivering near-diffraction-limited imaging become available.

By slicing the relatively wide slit matched to natural or partially-corrected seeing, the size of spectrographs can be reduced since the required beam diameter scales inversely with the slicing factor for fixed resolving power. However, other effects act against this trend, placing limits on the amount by which the instrument volume can be reduced. This has taken careful account of in a simple spectrograph model that includes scaling parameters which we set to give a good fit to three different types of spectrograph used currently on 8-m telescopes. The model allows us to explore the range of fitting parameters consistent with the calibrators.

We adopt 4 different models, based on a pair of binary options. The first specifies whether the extra slices are accommodated within the same instrument or by multiple replicated spectrographs designed for a single slice. The second option specifies whether the camera focal ratio or the detector oversampling is modified to provide correct image sampling, subject to limitations imposed by the Nyquist citerion and the fastest practicable camera design. The cost of the instrument is estimated , following industry practice, on the volume of the instrument and by scaling the detector sub-system independently by the number of pixels.

The main conclusions are as follows.

- Slicing can potentially reduce the cost of an instrument by large factors: 2-8 for medium spectral resolution to at least 70 for high resolution spectrographs.

- The best option is to include the slices within a single spectrograph rather than distribute them to replicas.

- For visible/near-infrared multiplexed spectrographs working at low-medium spectral resolution ($R = 5000$) a slicing factor of about 5 is optimal

- For a blue/visible spectrograph working at high resolution (R=150,000), the optimum slicing factor is 10-20.

- These conclusions assume a harsh scaling law for the detector sub-system. Relaxing this produces a substantial reduction in cost for large slicing factors, making the case for slicing more compelling.

- There is an upgrade path to integral field spectroscopy as the quality of the input images improves simply by modifying the data reduction software. However this may require larger slicing factors than the cost-optimised values proposed above, thereby requiring a larger initial cost.

Finally, this model may be applied to many other instrument concepts: only a very basic, and rather conservative, set has been examined here. The author will be happy to discuss this further with interested people.